\documentclass[sigconf]{acmart}

\setcopyright{none}
\renewcommand\footnotetextcopyrightpermission[1]{}
\usepackage[utf8]{inputenc}
\usepackage{color}

\AtBeginDocument{%
  \providecommand\BibTeX{{%
    \normalfont B\kern-0.5em{\scshape i\kern-0.25em b}\kern-0.8em\TeX}}}

\begin{document}

\title{Performance of Devito on HPC-Optimised ARM Processors}

\author{Hermes Senger}
\affiliation{%
  \institution{Universidade Federal de São Carlos}
  \city{São Carlos}
  \state{SP, Brazil}
}
\email{hermes@ufscar.br}

\author{Jaime F. de Souza}
\affiliation{%
  \institution{Universidade Federal de São Carlos}
  \city{São Carlos - SP}
  \country{Brazil}}

\email{jaimefreire.souza@gmail.com}

\author{Edson S. Gomi}
\affiliation{%
  \institution{Universidade de São Paulo}
  \city{São Paulo}
  \country{Brazil}
}

\author{Fabio Luporini}
\affiliation{%
  \institution{Imperial College London}
  \city{London}
\country{UK}}

\author{Gerard J. Gorman}
\affiliation{%
 \institution{Imperial College London}
 \city{London}
\country{UK}}

\maketitle


We evaluate the performance of Devito, a domain specific language (DSL) for finite differences on Arm ThunderX2 processors.
Experiments with two common seismic computational kernels demonstrate that Arm processors can deliver competitive performance
compared to other Intel Xeon processors.


\section{HPC-Optimised ARM Processors}
\label{sec:intro}

Arm processors such as the Huawei (Kunpeng 920), Ampere {(eMAG)}, Fujitsu (A64FX), and Marvell (ThunderX2) are emerging as an alternative
to traditional x86 architectures for HPC. The Isambard \cite{Mcintosh-Smith2018a} is the largest Arm based HPC production
system in Europe, and the first Cray XC50 (Scout) system to combine Arm based processors (32-core Marvell ThunderX2) with Cray's Aries
interconnect. Each of the 42 blades integrates 4 nodes with two 32-core Marvell ThunderX2 CPUs with 256 GB of DDR4 DRAM. The whole system
has 10,752 Armv8 cores. Recent studies compared the single node performance and multi-node scalability of Arm systems \cite{Mcintosh-Smith2018a, McIntoshSmith2019, zhang2011stable}. They demonstrated that for a wide range of applications, an Arm based supercomputer provides levels of performance competitive with state-of-the-art HPC-optimized
processors (e.g. Intel Skylake and Broadwell) with very attractive performance per dollar ratio.

\section{Devito - A DSL for finite differences}

Devito is a DSL and a framework for the solution of PDEs based on the finite difference method (FDM) \footnote{
\url{https://www.devitoproject.org/}}. Initially designed to implement high-performance wave propagation solvers and adjoint-state methods 
for seismic imaging problems, Devito allows concise expression of FDM and general stencil operations symbolically. Devito uses SymPy for the
generation and manipulation of stencil expressions and a pipeline of compilers and libraries to automate code generation, by applying several
symbolic, and loop optimisation to generate highly efficient implementations of algorithms for different hardware architectures \cite{Luporini2018}.
Originally, Devito was designed to support Intel Xeon and Intel Xeon Phi, and early investigation on different optimisation strategies
which had not been considered by other stencil compilers.
For example, most stencil compilers focus on cache reuse optimisation, while stencils like TTI \cite{zhang2011stable}
have very high arithmetic intensity, which results in elevated register pressure and requires specific optimisation techniques \cite{Luporini2018}.
In addition, there are mathematical operators that fall outside the regular stencil programming model but need to be supported for practical
applications. For example, source injection, interpolation at receivers and complex boundary conditions rely upon computation that is both
sparse and irregular. Currently, parallelism is supported by OpenMP and MPI which are integrated to the Devito 
stack.

\begin{table}[h]
  \caption{ Hardware specifications (per socket). 
  Intel processors can slow down clock when all cores execute AVX512.}
  \label{tab:specs}
  \begin{tabular}{rccc}
  
    \toprule
                      & ThunderX2 & Intel 5120 & Intel 6126 \\
    \midrule
    Cores/socket  & 32        & 14        & 12 \\
    Threads/core     &1,2 or {\bf 4}& 2       & 2 \\
    Clock frequency (GHz)  & 2.2     & 2.2/{\bf 1.6}    & 2.6/{\bf 2.3}   \\
    Vector unit width& 128-bit   & 512-bit   & 512-bit \\
    \midrule
    Max. \# FP64 ops./cycle& 8          & 16        & 32  \\
    \midrule    
    Max. perf.(FP64) Gflop/s & 563.2    & 358.4     & 883.2 \\
    Linpack (FP64) Gflop/s  & 410.5  &   345.4 &  695.0 
    \\
    \midrule        
    L1 cache (core)   & 32 KB    & 32 KB     & 32 KB \\
    L2 cache (core)   & 256 KB   & 1 MB      & 1 MB  \\
    L3 cache (shared) & 32 MB    & 19 MB     & 19 MB \\
    \midrule
    Memory     & 128 GB   & 192 GB    & 96 GB \\
    Memory channels   & 8        &  6        &  6 \\
    Maximum  bandwidth & 160.0 GB/s & 107.3 GB/s & 119.2 GB/s \\
    Measured bandwidth  & 118.7 GB/s &  64.5 GB/s &  70.9 GB/s \\
  \bottomrule
\end{tabular}
\end{table}

\section{Performance Evaluation}

We experimented on the Isambard system, which was described in Section \ref{sec:intro}. Single socket performance is compared against an Intel Xeon Gold 5120
and an Intel Xeon Gold 6126. See Table \ref{tab:specs} for specifications of all three processors. Memory bandwidth was measured with STREAM benchmark  compiled with GCC for Intel processors. For the Arm we used CCE which presented slightly better results. All experiments were executed 10 times and the best bandwidth was considered.
To evaluate the performance of Devito, we used two benchmarks: ($i$) the {\it acoustic} wave equation which models the propagation of an isotropic acoustic wave; and ($ii$) the Tilted Transverse Isotropy ({\it TTI}) model \cite{zhang2011stable}, which is a representative of state-of-art wave propagators for seismic imaging in production codes today. The full model specification, its finite difference schemes, and implementation using Devito are presented in \cite{louboutin2017performance}.
For the experiments, both {\it TTI} and {\it acoustic} equations are discretized with second-order in time and varying space orders of 4, 8, 12 and 16. 
The experiments for the two velocity models use $512^3$, $768^3$, and $1024^3$ grid points with a grid spacing of 20 m. The time step is modelled
for one second, resulting in 327 time steps for {\it acoustic}, and 415 time steps for {\it TTI}. Devito implements a lowering process (from mathematical equations down to C++ code), and integrates with other compilers. For experiments on Arm, we used Devito 3.4, Cray compiler 8.7.9, and GCC 8.2. For Intel processors, we used GCC 7.4. For the sake of reproducibility, we used only one thread per physical core and threads pinning to the processor cores.

The first experiment measures the performance of an increasing number of threads running on a single socket. ThunderX2 presents competitive execution times for both benchmarks, compared to the Intel Xeon 5120 and the Intel Xeon 6126 (Table \ref{tab:scalingCol}). While the single thread performance is better on the Xeon than the ThunderX2, the ThunderX2 delivers competitive performance to the Xeon when all
cores are utilised. This is due to the fact that the benchmark is memory bound (low operational intensity) and the ThunderX2 has a much higher
memory bandwidth than the Xeon's. 

\begin{table}[ht]
  \caption{The execution times, speedup (S), and the amount of GFLOPS (G) on a single socket, $512^3$ grid, and SO=4.}
  \label{tab:scalingCol}
    \begin{tabular}{  c | c  c  c | c c c }
    \hline
    \hline
      \multicolumn{7}{c}{\bf ThunderX2} \\ \hline
     & \multicolumn{3}{  c |}{\bf acoustic} & \multicolumn{3}{c}{\bf TTI}  \\
    {\bf T} & time & S  & G  & time & S  & G   \\
     \hline
        1 & 288.1  & 1.0  & 6.0  & 1879.4 & 1.0 & 7.6 \\ 
        2  & 145.2 & 2.0  & 11.9  & 977.1 & 1.9 & 14.6  \\
        4  & 73.5 & 3.9  & 23.5  & 489.1 & 3.8 & 29.2   \\ 
        8  & 37.6  & 7.7  & 46.0  & 245.5 & 7.7 & 58.2  \\ 
        16 & 19.3 & 14.9 & 89.7  & 126.3 & 14.9 & 113.1 \\
        32 & 11.0 & 26.3 & 157.7  & 67.3 & 27.9 & 212.3 \\
  \hline
  \hline
     \multicolumn{7}{c}{\bf Xeon Gold 5120} \\ \hline
     & \multicolumn{3}{| c |}{\bf acoustic} & \multicolumn{3}{c}{\bf TTI}  \\
    {\bf T} & time & S  & G  & time & S  & G \\   \midrule
        1 & 93.5 & 1.0  & 18.5  & 573.0 & 1.0 & 24.9 \\ 
        2  & 46.9 & 2.0  & 36.9  & 277.2 & 2.1 & 51.5 \\  
        4  & 25.2 & 3.7  & 68.7  & 162.2 & 3.5 & 88.1 \\  
        8  & 16.8  & 5.5 & 102.6  & 102.1 & 5.6 & 139.9 \\  
        14 & 14.7 & 6.4 & 117.7  & 69.2 & 8.3 & 206.5 \\  
        \hline
        \hline
    \multicolumn{7}{c}{\bf Xeon Gold 6128} \\ \hline
    & \multicolumn{3}{| c |}{\bf acoustic} & \multicolumn{3}{c}{\bf TTI} \\
    {\bf T} & time & S  & G  & time & S  & G \\   \midrule
        1 & 85.9 & 1.0  & 20.1 & 505.3 & 1.0 & 28.3 \\  
        2  & 42.6 & 2.0  & 40.5  & 246.3 & 2.1 & 58.0 \\   
        4  & 23.0 & 3.7  & 75.2  & 131.4 & 3.8 & 108.7   
        \\ 
        8  & 14.9  & 5.8 & 116.2  & 81.3 & 6.2 & 175.7    
        \\ 
        12 & 13.6 & 6.3 & 126.9  & 61.4 & 8.2 & 232.6    
        \\
        \hline
        \hline
\end{tabular}
\end{table}

\begin{figure}[h]
  \centering
  \includegraphics[width=\linewidth]{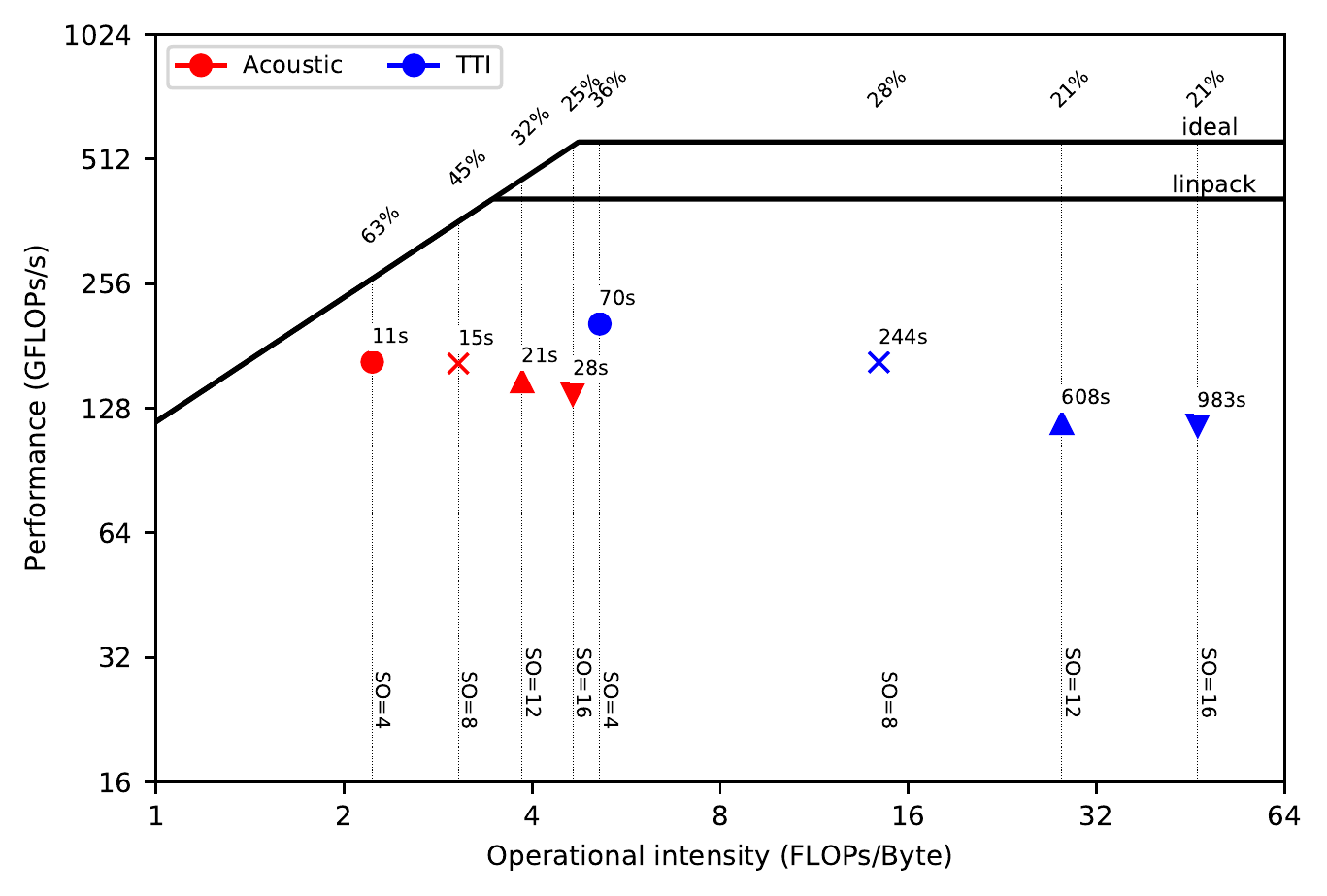}
  \caption{Arm ThunderX2, $512^3$ grid points.}
  \label{roofline-arm-acoustic}
\end{figure}

The next experiment measured the performance of the code generated by Devito in terms of the maximum performance for the Arm processor (in Fig. \ref{roofline-arm-acoustic}).
We performed a complete set of experiments including two simulation models (acoustic isotropic, TTI), two compilers (GCC-8, and CCE),
three Devito optimisation modes (basic, aggressive, DSE), three grid sizes ($512^3$, $768^3$, and $1024^3$ points), and 20m steps.
For the GCC we used the flags {\tt  -O3 -g -fPIC  -march=native  --fast-math -shared -fopenmp}, and for the CCE compiler we used the flags {\tt -O3 -g -fPIC -shared -homp}. 
The results shown for the Arm processor were produced by GCC, which presented slightly better performance compared to CCE. In total, 288 simulations were executed being replicated three times and averaged. The variance observed is negligible (< 1\%).


\section{Findings}

The results presented demonstrate that Arm based processors are capable of delivering performance similar to state-of-the-art Intel Xeon processors
for the execution of seismic inverse problems. Additionally, Devito is shown to be capable of generating efficient high performance code for Arm processor. All models compiled and ran successfully, and no architecture specific code tuning was necessary to achieve high performance.

\begin{acks}

This research was carried out in association with the ongoing R\&D project registered as ANP 20714-2 - ``Desenvolvimento de técnicas numéricas
e software para problemas de inversão com aplicações em processamento sísmico'' (USP/Shell Brasil/ANP), sponsored by Shell Brasil under the
ANP R\&D levy as ``Compromisso de Investimentos com Pesquisa e Desenvolvimento''. This work used the Isambard UK National Tier-2 HPC Service
(http://gw4.ac.uk/isambard/) operated by GW4 and the UK Met Office, and funded by EPSRC (EP/P020224/1). Hermes Senger also thanks FAPESP 
(Contract number 2018/00452-2) for their support. 
\end{acks}

\bibliographystyle{ACM-Reference-Format}
\bibliography{sample-base}




\end{document}